# Combined flux compression and plasma opening switch on the Saturn pulsed power generator

Franklin S. Felber,[1] Eduardo M. Waisman,[2] and Michael G. Mazarakis[3]
[1]*Physics Division, Starmark, Inc., P. O. Box 270710, San Diego, California 92198, USA*
[2]*Sandia National Laboratories, Albuquerque, NM 87185, USA*

A wire-array flux-compression cartridge installed on Sandia's Saturn pulsed power generator doubled the current into a 3-nH load to 6 MA and halved its rise time to 100 ns. The current into the load, however, was unexpectedly delayed by almost 1 μs. Estimates of a plasma flow switch acting as a long-conduction-time opening switch are consistent with key features of the power compression. The results suggest that microsecond-conduction-time plasma flow switches can be combined with flux compression both to amplify currents and to sharpen pulse rise times in pulsed power drivers.

This paper suggests a simple explanation for certain unusual features observed in magnetic-flux-compression experiments on the Saturn pulsed power generator at Sandia National Laboratories. The goal of the experiments was to develop a flux-compression cartridge that, when installed on the Saturn generator, would amplify the current and sharpen the pulse rise time. The goal was achieved, but in an unexpected way. The amplified and sharpened current pulse was observed in the load only long after the Saturn pulse had peaked.

This paper briefly reviews the design of the flux-compression cartridges and the results of the experiments, presented in [1]. Then it suggests a picture of a plasma opening switch (POS) based on a plasma mass accelerated through a coaxial channel from the cartridge to the load. With this picture, the near-microsecond-long conduction time observed in the experiments is consistent with the amplified current and sharpened pulse rise time measured in the load.

Flux-compression and inductive-storage technologies, including POS technology, have value for applications of high-power pulsed-power drivers as sources of x-ray radiation for indirect drive of inertial confinement fusion (ICF) capsules. Major cost savings on pulsed-power radiation drivers, such as a factor of a few or more, can potentially be achieved by innovative approaches to power compression, such as magnetic-flux compression. Long-rise-time generators operate at lower voltages and cost less. Since magnetic-flux compression can potentially produce savings in a national investment in Z-pinch-driven ICF of hundreds of millions of dollars per facility at the scale of energy production, great efforts have been expended in many countries over the past decades to develop and improve flux-compression and inductive-storage technologies.

For over half a century, armatures driven by high explosives (HE) were used for magnetic flux compression. HE-driven magnetocumulative generators (MCGs) were fast [2] or efficient [3]. But even fast HE-driven MCGs are too slow for many applications. For example, a 1-m-long optimized coaxial tapered sweeping-wave MCG, such as was developed in [4], could not achieve a final current-doubling time less than about 10 μs.

Plasma speeds in certain configurations can be substantially faster than explosive detonation velocities. The first concepts for using plasmas to compress magnetic flux density proposed implosion of liners by laser ablation [5] and by Z pinches [6]. The Z-pinch method compressed a 100-kG seed field to 42 MG on the Proto II pulsed power driver [7]. The laser-ablation method compressed a >50-kG seed field to 30 to 40 MG on the OMEGA laser [8].

The first concept to use plasmas to compress magnetic flux density for pulse sharpening and current amplification in pulsed power drivers was proposed by Léon *et al.* [9]. The concept is to use the primary current of a pulsed power driver to drive the magnetic flux of a secondary current from a large volume with high inductance into a smaller volume with lower inductance. Since electrically-driven plasma velocities in pulsed power drivers can exceed 10 cm/μs, and plasma armatures may only need to move a few cm or less, the compression might be complete in hundreds of ns, with final current-doubling times of 100 ns or less.

Before the Saturn experiments, the flux-compression concept proposed in [9] was tested experimentally on Z at Sandia, ECF at Centre d'Etudes de Gramat in France, and Hawk at the Naval Research Laboratory. On Z, primary currents between 15 and 18 MA amplified secondary currents to about 20 MA in a 0.1-nH load [10, 11]. On ECF, a primary current of 2 MA amplified the secondary current to about 3 MA, but it is not clear whether this amplified current was measured in an inductive load or was embedded in the armature plasma [10 – 12]. On Hawk, flux-compression cartridges amplified the 760-kA current up to 940 kA in a 1-nH load [13].

K-shell yields from 100-ns to 200-ns implosions of large-diameter structured *gas-puff* Z-pinch loads were produced at the Decade Radiation Test Facility near the theoretical maximum for that class of pulsed power generators, including Saturn, and seem to have limited room for improvement [14,15]. Flux compression, however, offers the possibility of substantial improvements of K-shell yields of *wire-array* plasma radiation source (PRS) loads, such as those made of titanium or copper, and could even be an enabling technology for high-Z K-shell radiation from small-diameter loads. Current amplification by flux compression might also improve the yield of ICF targets compressed isentropically by soft x-ray radiation from nested wire arrays [16].

The flux-compression cartridges developed and tested in [1] were based on stable 'skinny-armature' cartridge designs proposed earlier [17] and tested on Hawk [13]. Long 'skinny armatures' have greater inductance and less instability growth. 'Skinny-armature' flux-compression cartridges were installed on Saturn to amplify and sharpen the current into a 3-nH static inductive load. These 18-cm-long cartridges were recessed within Saturn to retain Saturn's diagnostic and radiation lines of sight to the load region.

In the double-post-hole-convolute, long-pulse mode, Saturn typically delivers about 6 MA to a load. Since no other appropriate source of seed magnetic flux was available, Saturn was made to produce two relatively independent currents by removing the posts convoluting the upper pair of magnetically insulated transmission lines (MITLs) from the lower pair. This *ad hoc* adaptation doubled the 9- to 10-nH machine inductance and halved the maximum current in a pair of MITLs to 3 MA.

In most shots, six of the 36 independent pulsed-power modules provided the secondary current through the lower pair of MITLs. The remaining modules provided the primary current, which drove the imploding wire-array-plasma armature that compressed the seed magnetic flux. Attempts to fire the secondary-current modules before the primary current on some shots were unsuccessful. All three shots in which all 36 modules were triggered simultaneously in the 200-ns long-pulse mode showed the same microsecond-conduction-time behavior. A major modification to the Saturn middle anode made Shot 7 the most successful of this series.

Figure 1(a) is a schematic illustration of the flux-compression cartridge design used in Shot 7. The initial inductance between the armature and stator was about 30 nH, with another 1.3 nH in the neck (channel) to the load and 1.7 nH in the load volume itself. The flux density was to have been compressed from a 33-nH volume into about a 3-nH volume. In the face of all possible loss mechanisms, including diffusion, shocks, and flux-trapping against the stator by Rayleigh-Taylor bubbles and plasma precursors, the aim was for the load current to be double the primary current. Although this aim was achieved, our circuit models did not even roughly simulate the largely unexpected current waveforms.

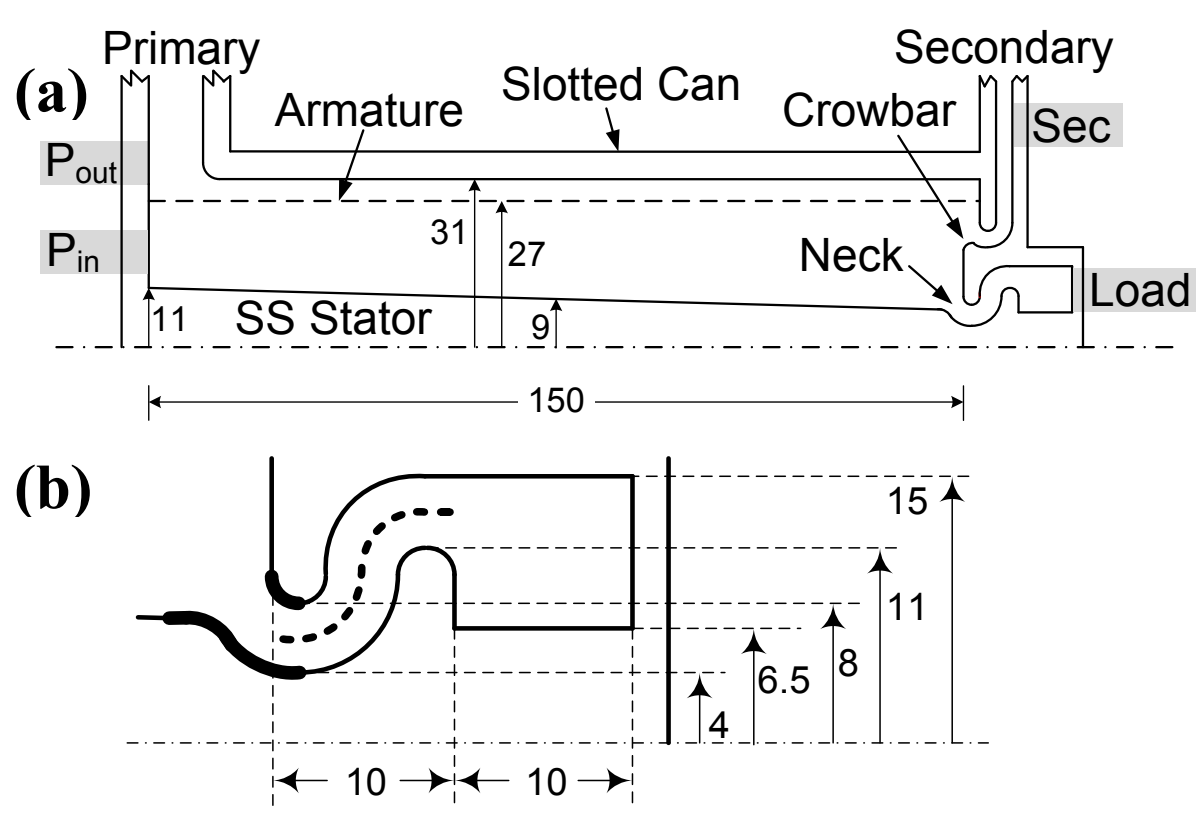

FIG. 1. Flux-compression cartridge: (a) sideways, b-dot locations shaded; (b) neck and load region, plasma path (dashed), 5-cm$^2$ area (heavy solid). All dimensions in mm. Adapted from [1].

Besides post-mortem analysis of hardware, b-dot probes were the only diagnostic in all shots. Sandia calibrated the Sandia b-dot probes *in situ* after installation on the cartridges, and performed the data acquisition and processing, with two oscilloscope channels per probe. In addition to the usual Saturn machine b-dot probes, six pairs of b-dot probes were used on each shot. Each probe of a pair was located on the side of the cartridge opposite the other and had the opposite polarity. For clarity, all probe data is displayed in Fig. 2 with positive polarity.

As shown in Fig. 1(a), the pair of '$P_{out}$' probes measured the primary current at all times. The pair of '$P_{in}$' probes measured the secondary current inside the cartridge before the armature passed, and the primary current (with opposite polarity) after it passed. Two pairs of 'Sec' probes at different radii measured the secondary current outside the cartridge. One pair of 'Load' probes was of the type normally used on the Z machine; the other, normally used on Saturn. Only the load-current data from Sandia's Z-type probes are presented in Fig. 2, because the Z-type probes were designed for more adverse environments and gave cleaner, less noisy results.

In Shot 7, the Saturn peak primary current (measured at '$P_{out}$') of 3.1 ± 0.3 MA was doubled to 6.3 ± 1.0 MA

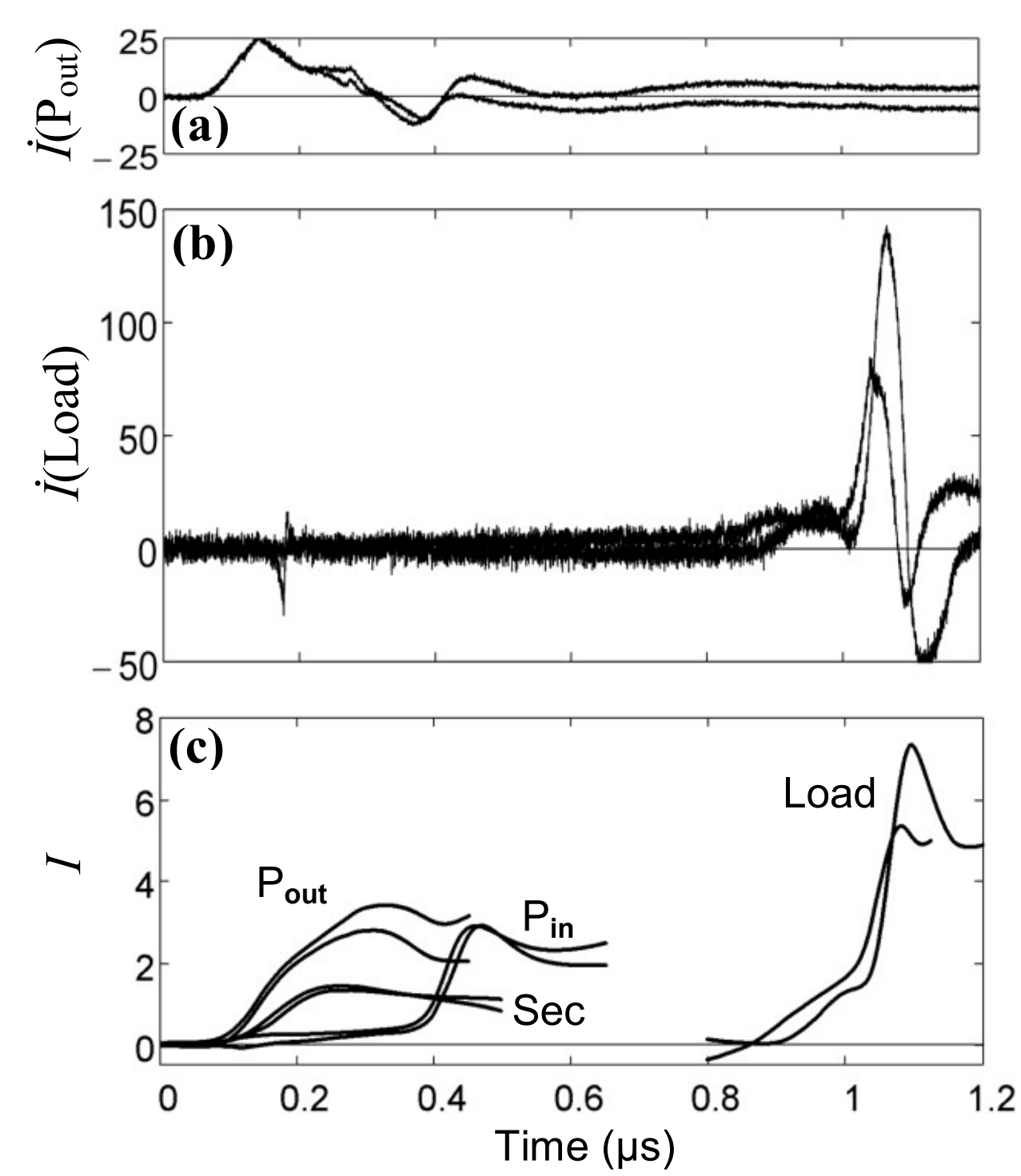

FIG. 2. B-dot data vs. time (in μs) on Shot 7: (a) rates of change $\dot{I}$ (in MA/μs) of primary ($P_{out}$) currents and (b) load currents; and (c) primary ($P_{out}$ and $P_{in}$), secondary (Sec), and load currents (in MA). Adapted from [1].

in the load, and the 200-ns rise time of the Saturn primary current was halved to about 100 ns in the load. The peak rate of change of the primary current of 26.1 ± 0.05 MA/μs was increased to 114 ± 29 MA/μs in the load after almost a 1-μs delay.

The secondary current, apparently unnecessary for Shot 7, was shorted out before it got inside the cartridge. By the time the armature reached the '$P_{in}$' probe, this probe measured 300 kA circulating between the armature and stator, probably from diffusion through the armature plasma. As the armature passed over the '$P_{in}$' probe, the current rapidly rose to the 3-MA level of the primary current in '$P_{out}$' , as expected. That the '$P_{in}$' current had the same polarity as the '$P_{out}$' current before the armature passed the '$P_{in}$' probe, suggests the 300-kA '$P_{in}$' current initially was diffused through the armature, rather than from the secondary current. That the 300-kA '$P_{in}$' current built up slowly and smoothly, also suggests the source was diffusion rather than, say, convection by precursors.

Until late in the shot, the load also seems to have been shorted out of the circuit, probably by plasma precursors at the entrance to the neck. A transient 100-kA pulse, seen (unintegrated) in Fig. 2(b), was detected by both Z-type b-dot probes in the load. After that, current in the load was not noticeable until almost 800 ns after the primary and secondary currents began. (Without correcting for the drift of the integrated probe data by zeroing the load current at that point, the peak load current would have averaged 6.7 MA, instead of 6.3 MA.)

A relatively long conduction time followed by a sharp current rise in a load is characteristic of an inductive-storage circuit with a current-interrupting switch. An early concept of a plasma opening switch based on a current penetrating axially through a plasma channel at the front of a magnetoresistive wave was presented in [18]. A related concept is the plasma erosion opening switch (PEOS), in which a plasma is injected into a channel, conducting the current as a short circuit, and opening the circuit when a current threshold is exceeded [19–21]. Experiments in the mid-1980's demonstrated the potential for plasma opening switches to operate with conduction times as long as 1 μs [22–25].

Plasma flow switches (PFSs), which open when a plasma reaches the end of a channel, were reviewed in [26] and modeled in [27]. A typical PFS comprises a wire array and a barrier foil in a coaxial channel upstream from the load. The plasma initiated at the wire array is assembled at the barrier foil before being accelerated by the $J \times B$ force along the channel to the load.

Conditions for a microsecond-conduction-time PFS may have been created naturally in the flux-compression cartridges in the three shots in which switch-like behavior was observed. The model of the PFS in our flux-compression experiments assumes a diffuse plasma mass shorting the gap at the entrance of the neck in Fig. 1(b). The neck serves as a coaxial channel from the flux-compression cartridge to the load. The $J \times B$ force quickly snowplows the diffuse plasma into a dense, highly conductive sheath, which is then accelerated by the magnetic pressure along the channel to the load during the conduction time of the opening switch.

As the plasma exits the channel, the magnetic pressure is stronger in the radial direction than it is on average in the axial direction, and there should be much less trailing plasma keeping contact with the channel muzzle than there is in the bulk plasma moving axially. It is therefore reasonable to suppose that the bulk plasma cannot travel much beyond a channel width before the magnetic field penetrates radially through the trailing plasma to open the switch. (Alternatively, the plasma might convect or 'commutate' the current into the walls of the load in a time of order of the plasma thickness divided by its speed.)

In the cartridge used for Shot 7, a 30-mg plasma mass driven by a 6.3-MA current would have an acceleration of 6.3 cm/μs$^2$ and would transit the 1.5-cm length of the channel in 740 ns. When the plasma exits the channel, its velocity would be 4.0 cm/μs, and it would move a channel width in about 100 ns.

In the Shot 7 data shown in Fig. 2(c), the time elapsed from the peak of the primary current to the current in the load is about 700 to 800 ns, corresponding to the plasma transit time of the channel and the conduction time of the PFS estimated above. The rise time of the load current in Fig. 2(c) is about 100 ns, corresponding to the opening time of the PFS estimated above.

These estimates agree with Shot 7 if about 30 mg of plasma initially conducts current at the entrance of the neck. Less than 5 cm$^2$ of Type "C" parylene coating ablated there could have provided this 30 mg of plasma. A 5-cm$^2$ coated area is indicated by the thick curves in Fig. 1(b).

These estimates suggest that a PFS in a single stage can be combined with flux compression, first to amplify the current, and then to sharpen the rise time of the amplified pulse to the 100 ns or less needed for a PRS from pulse durations as long as 1 to 2 μs. In the embodiment shown in Fig. 3, the flux compressor can amplify the current of a slow generator and the PFS can sharpen the pulse potentially by much more than the factor of 4 increase in $\dot{I}$ seen in Fig. 2.

The PFS could be any plasma with low diffusivity that will maintain its integrity as it is accelerated by magnetic pressure through a long coaxial channel. Under some conditions a bare wire array might even be used without a barrier foil since, with the circuitous channel design shown in Fig. 3, plasma precursors from the wire array would not flow directly into the load.

The channel might also include a plasma-boundary-layer trap to improve PFS performance [27]. The PFS might even be made to open by mechanisms more close-ly resembling those of a PEOS than a coaxial plasma gun. For example, plasma can be eroded by being exhausted through apertures in the channel or entrained against the channel walls. Although opening the switch by erosion while the plasma is still within the channel has some advantages, opening upon exiting the channel might be more effective in some cases, such as if the channel cannot be designed long enough for sufficient erosion to occur.

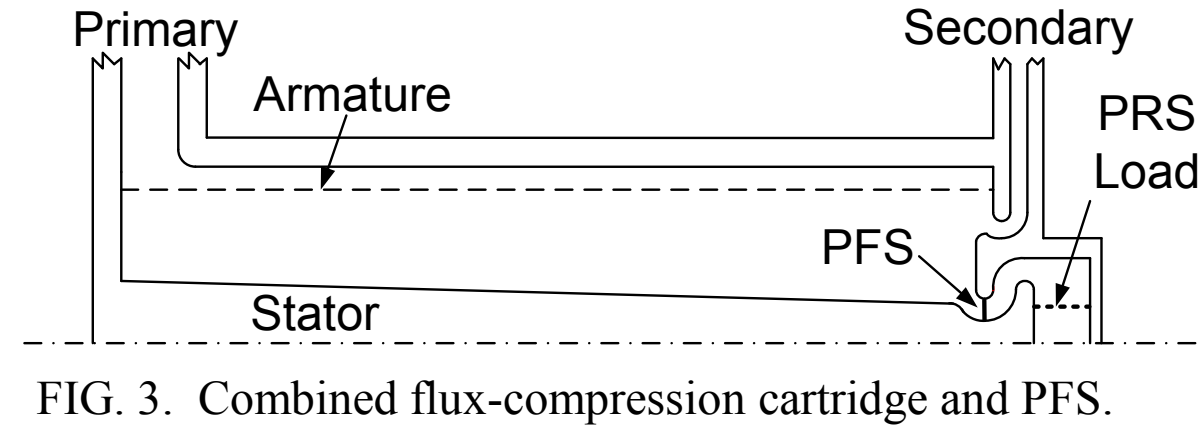

FIG. 3. Combined flux-compression cartridge and PFS.

The conduction time and power compression that can be achieved with a PFS are limited by: (i) the requirement for the channel transit time to be shorter than the diffusion time; (ii) the number of Rayleigh-Taylor e-foldings, which is proportional to the square root of the channel length; (iii) inductive energy lost to kinetic energy and heating of the plasma; and (iv) parasitic inductance of the channel.

Numerous practical advantages might result from inserting a PFS or its equivalent near a channel entrance, as in Fig. 3, including: (i) good shot-to-shot reproducibility; (ii) control of the conduction time and opening time; (iii) reduced entrainment and loss of magnetic flux; (iv) protection of the PRS load from the long voltage pulse; and (v) compatibility with flux compression.

In summary, wire-array cartridges were installed on Saturn to amplify the current and sharpen the pulse by flux compression. Unexpectedly long delays of almost 1 μs were observed between the Saturn pulses and the currents into the loads on three shots, including one shot in which the load current was doubled and its rise time halved. The long delay and the sharp pulse rise time are consistent with a naturally occurring PFS, which opened only after a plasma was ejected from the channel into the load volume. The results suggest that microsecond-conduction-time PFSs might be particularly well-suited to be combined with flux compression in amplifying currents and sharpening pulse widths in existing and next-generation pulsed power drivers. If so, the results could provide a significant step towards the long-term goal of applying flux compression to achieving substantial savings in the capital cost of next-generation facilities for radiation simulation and for energy production by Z-pinch-driven ICF.

1. F. Felber *et al*., Conf. Record, IEEE/NPSS Int. Conf. on Plasma Sci., San Diego, 2009, Paper IOC3-3.
2. J. W. Shearer *et al*., J. Appl. Phys. **39**, 2102 (1968).
3. A. D. Sakharov *et al*., Sov. Phys. – Doklady **10**, 1045 (1966).
4. F. S. Felber *et al*., in *Ultrahigh Magnetic Fields*, V. M. Titov and G. A. Shvetsov, eds. (Nauka, Moscow, 1984), p. 321.
5. M. A. Liberman and A. L. Velikovich, J. Plasma Phys. **31**, 381 (1984); G. D. Bogomolov, A. L. Velikovich, and M. A. Liberman, in *Ultrahigh Magnetic Fields*, V. M. Titov and G. A. Shvetsov, eds. (Nauka, Moscow, 1984), p. 232.
6. F. S. Felber, M. A. Liberman, and A. L. Velikovich, Appl. Phys. Lett. **46**, 1042 (1985).
7. F. S. Felber *et al*., Phys. Fluids **31**, 2053 (1988).
8. O. V. Gotchev *et al*., Phys. Rev. Lett. **103**, 215004 (2009).
9. J. F. Léon *et al*., Proc. 12th IEEE Int. Pulsed Power Conf., 1999 (IEEE, Piscataway, NJ, 1999), p. 275.
10. M. Bavay *et al*., Conf. Record, IEEE Int. Conf. on Plasma Sci., Banff (IEEE, Piscataway, NJ, 2002), p. 212.
11. M. Bavay, Doctor of Sciences Thesis, Université de Paris sud UFR scientifique d'Orsay, 8 July 2002.
12. P. L'Eplattenier *et al*., 5th Int. Conf. on Dense Z-Pinches, AIP Conf. Proc. **651**, 51 (2002).
13. F. Felber *et al*., Conf. Record, IEEE Int. Conf. on Plasma Sci., Monterey, 2005, p. 183.
14. P. L. Coleman *et al*., 6th Int. Conf. on Dense Z-Pinches, Oxford, 2005, AIP Conf. Proc. **808**, 163 (2006).
15. H. Sze *et al*., Phys. Rev. Lett. **95**, 105001 (2005).
16. M. Cuneo *et al*., Phys. Rev. Lett. **95**, 185001 (2005).
17. The 'skinny armature' concept first appeared in F. S. Felber and E. M. Waisman, Proc. Workshop on Innovative Power Amplification in Vacuum, Maxwell Physics Int'l., San Diego, 5–6 Sep. 2000.
18. F. S. Felber *et al*., Appl. Phys. Lett. **41**, 705 (1982).
19. B. V. Weber *et al*., IEEE Trans. Plasma Sci. **PS-15**, 635 (1987).
20. B. V. Weber *et al*., IEEE Trans. Plasma Sci. **19**, 757 (1991).
21. E. Waisman *et al*., SSS Report SSSDFR8910651, DTIC Accession No. ADA282701 (1987).
22. B. M. Koval'chuk and G. A. Mesyats, Sov. Phys. Dokl. **30**, 879 (1985).
23. D. D. Hinshelwood *et al*., Appl. Phys. Lett. **49**, 1635 (1986).
24. R. J. Commisso *et al*., Phys. Fluids B**4**, 2368 (1992).
25. D. Hinshelwood *et al*., Phys. Rev. Lett. **68**, 3567 (1992).
26. P. J. Turchi *et al*., IEEE Trans. Plasma Sci. **PS-15**, 747 (1987).
27. R. L. Bowers *et al*., LANL Technical Report LA-UR-93-0014, OSTI ID: 7012135 (1 Jan. 1992).